\documentstyle[preprint,aps]{revtex}
\tightenlines
\begin{document}
\draft
\title{Scaling of the Conductivity with Temperature and Uniaxial Stress 
in Si:B at the Metal-Insulator Transition.}
\author{S. Bogdanovich and M. P. Sarachik}
\address{Physics Department, City College of the City University of 
New York, New York, New York 10031}
\author{R. N. Bhatt}
\address{Department of Electrical Engineering, Princeton University, 
Princeton, New Jersey 08544-5263}
\date{\today}
\maketitle
\begin{abstract}
Using uniaxial stress, $S$, to tune Si:B through the metal-insulator 
transition at a critical value $S_c$, we find the dc conductivity 
at low temperatures shows an excellent fit to the scaling
form $\sigma (S, T) = AT^xf[(S-S_c)/T^y]$ on both sides of the transition.  
The scaling functions yield the conductivity in the 
metallic and insulating phases, and allow a reliable determination of the 
temperature dependence in the critical regions on both sides of the transition.
\end{abstract}
\pacs{PACS numbers: 71.30.+h}

The metal-insulator transition in doped semiconductors and amorphous metal-
semiconductor mixtures is a continuous 
quantum phase transition which occurs in the limit of 
zero temperature.  A scaling approach similar to that used for continuous phase 
transitions driven by temperature, within which the properties of 
the system do not depend on microscopic details and 
are controlled by a diverging length scale, suggests that the 
conductivity in the vicinity of the transition (critical) point
can be described by a scaling form:
\begin{equation}
\sigma (t, T) = \sigma_c (T) f[(t-t_c)/T^y].
\label{eq}
\end{equation}
where 
$t$ is the control parameter (such as dopant concentration, magnetic 
field or uniaxial stress) that drives the transition at the critical 
value $t = t_c$, and
\begin{equation}
\sigma_c(T) = A T^x
\label{eq}
\end{equation}
is the low-temperature limit of the conductivity at the critical point.
By identifying the diverging time scale $\tau$ at the transition with
$ \hbar /kT $, and assuming conventional dynamical scaling
$\tau \propto \xi^z$, where
$\xi (t) \propto (t-t_c)^{-\nu}$  is the diverging correlation 
length scale as the transition is approached, one easily obtains
$ y = 1/z\nu $, and $ x = \mu /z\nu $, where $\mu$ is the critical
exponent characterizing the onset of metallic conductivity at
zero temperature:
\begin{equation}
\sigma (t, T \rightarrow 0) \propto (t-t_c)^\mu.
\label{eq}
\end{equation}
The applicability of this scaling formulation\cite{dynamical} to the metal-insulator transition 
was first demonstrated for non-interacting electrons by
Wegner\cite{wegner} and by Abrahams et al.\cite
{gang}, and subsequently extended to incorporate electron-electron interactions 
by many, including McMillan\cite{mcmillan}, Finkelshtein\cite{finkelshtein}, 
Castellani {\it et al.}\cite{castellani}, and Belitz and Kirkpatrick\cite{bandk}.  
Within this theoretical framework, the values of the critical exponents 
$\mu$, $\nu$ and $z$ are determined by the symmetry of the effective 
field theory, and depend on the presence or absence of
symmetry-breaking fields, such as magnetic field,
spin-orbit interactions, or magnetic impurities, which determine 
the universality class of the system\cite{reviews,analysis}.

The problem of metal-insulator transitions has a venerable history\cite{mott}.
Data for amorphous metal-semiconductor mixtures\cite{bishop}, magnetic semiconductors
\cite{vonM}, and heavily compensated persistent
photoconductors,\cite{katsumoto}, all suggest 
an exponent $\mu \approx 1$.  In contrast, although the continuous nature of 
the transition was first demonstrated in uncompensated doped Si and despite 
considerable effort over more than two decades, a consensus regarding the 
critical behavior in uncompensated doped semiconductors has yet to emerge\cite{mottsarachik}.  
There continues to be debate concerning a number of 
fundamental issues. Thus, for example, (i) the value of the critical conductivity 
exponent $\mu$ has been variously cited as equal to 1/2 and 1; (ii) the breadth of the critical regime, and thus the range of the 
critical parameter t one can safely 
use to determine the exponent is not known; and (iii) the form of $\sigma_c (T)$ in the 
critical region very near the transition has been claimed to be 
$\propto T^{1/2}$ and $\propto T^{1/3}$\cite{T^1/3}.

The procedure generally used to determine the critical exponent $\mu$ 
entails measuring the conductivity to as low a temperature as possible to 
obtain a single extrapolated T=0 value to be used in the fit to Eq. (3).  In contrast, 
full scaling with temperature, Eq. (1), obviates the need for 
potentially unreliable extrapolations to T=0 and yields a determination of 
$\mu$ based on all data taken at all temperatures.  Inspection of published 
data for amorphous metal-semiconductor 
mixtures such as NbSi,\cite{bishop} and the persistent photoconductor Al$_{0.3}$
Ga$_{0.7}$As,\cite{katsumoto} indicates that the conductivity of 
these systems obeys Eq. (1) in the metallic phase\cite{parallel,rosenbaum}.
 On the other hand, the conductivity of Si:P measured by 
Paalanen {\it et al.}\cite{mikko} down to several 
mK does not obey scaling, while approximate temperature 
scaling has been reported by Stupp {\it et al.}\cite{hilbertcomment} on the metallic 
side very near the transition in the same system.

	In this paper we report conductivity measurements obtained using 
uniaxial stress to tune through the metal-insulator transition in Si:B.  As 
demonstrated by Paalanen {\it et al.}\cite{mikko} in their classic experiments in 
Si:P, uniaxial stress allows very fine control in a single sample, and gives 
precise relative determinations of the critical parameter $t$.  For a closely 
spaced set of stresses $S$ near the critical stress $S_c$, we demonstrate that 
the conductivity of Si:B obeys scaling, Eq. (1), on both sides of the 
metal-insulator transition.  We show that the conductivity in 
the critical regime is consistent with $\sigma_c \propto T^{1/2}$.  This 
is the same temperature dependence as has been calculated\cite{altandaron} 
and observed\cite{mikko} in the perturbative region on the metallic 
side (weakly disordered metal).  On the insulating side, we find that the conductivity crosses over to 
$\sigma \propto exp (T^*/T)^{1/2} $, the form expected for variable-range hopping in the presence 
of Coulomb interactions\cite{efrosshklovskii}, with a 
prefactor $\propto T^{1/2}$ corresponding to the temperature-dependence of 
the critical curve.

	A bar-shaped $8.0$x$1.25$x$0.3$ mm$^3$ sample of Si:B was cut with its long 
dimension along the [001] direction.  The dopant concentration, determined 
from the ratio of the resistivities\cite{bogdanovich} at 300 K and 4.2 K, 
was $4.84$x$10^{18}$ cm$^{-3}$.  
Electrical contact was made along four thin boron-implanted strips.  Uniaxial 
compression was applied 
to the sample along the long [001] direction using a pressure cell described 
elsewhere\cite{bogdanovich}.  Four-terminal measurements were taken at 13 Hz (equivalent 
to DC) for different fixed values of uniaxial stress at temperatures between 
0.05 and 0.76 K.  Measurements were restricted to the linear 
region of the I-V curves.

	Si:B is considerably more sensitive to stress than Si:P\cite{pollak}.  
This is because the acceptor state in Si:B has a four-fold degeneracy in the
unstressed cubic phase, which is lifted by uniaxial stress into two doublets,
each retaining only the Kramers degeneracy. By contrast, the six-fold valley degeneracy
(on top of the required Kramers or spin degeneracy)
of an effective mass donor in Si has already
been removed (even in zero stress) by the central-cell correction of the 
phosphorus dopants\cite{bhatt}.
A consequence of the additional degeneracy in Si:B is that uniaxial
compressive stress drives metallic Si:B into the insulating phase, unlike
Si:P where (relatively larger) stresses drive an insulating sample through
the transition into the metallic phase. This is in qualitative agreement with
predictions for effective mass donor systems
which take into account mass anisotropy\cite{bhatt2} as well as degeneracy
in the presence of electron correlation\cite{bhattandsachdev}.

The conductivity of Si:B measured at 4.2 K was found 
to be a linear function of stress.  Since the conductivity at 4.2 K is also 
approximately linear with dopant concentration for concentrations near the 
transition, it follows that $(S-S_c) \propto (n-n_c)$, and either may be used 
as the control parameter $(t-t_c)$ in the scaling equations (1) and (3).  
It has generally been assumed that the same value of the 
critical conductivity exponent $\mu$ should then be obtained by varying 
stress or dopant concentration.

	The conductivity as a function of temperature is shown on a log-log 
scale in Fig. 1 for twenty selected values of the uniaxial stress (data 
were taken for other stress values not shown).  The upper 
curves bend upward as the temperature is decreased, tending toward finite 
(metallic) conductivities at T=0, while the lower curves are concave
downward, indicating that they are in the insulating phase.  On a log-log 
scale the critical curve 
at $S=S_c$ is a straight line heading toward $\sigma = 0$ at $T=0$, and 
follows a power-law.

  Fits to full S-T scaling, Eq. (1), were 
carried out for different choices of $S_c$ ranging between 560 and 726 bar.  
Figure 2 shows a scaled plot of $\sigma /\sigma_c$ versus $T/(\Delta S/S_c)^{0.31}$ 
(where $\Delta S=S-S_c$) for the best choice, $S_c = 613$ bar, yielding 
exponents $x = 0.5$ and $y = 0.31$. The conductivity at the critical point $S=S_c$ 
thus exhibits a square root
dependence on temperature. A dependence
$\propto T^{1/3}$, claimed for several other systems at the critical point $t = t_c$,\cite{T^1/3} is 
decidedly inconsistent with our data in stressed Si:B.  In terms of the standard exponents, 
one obtains
$\mu = 1.6$ and $z\nu = 3.2$.  If we further assume (as is generally done)
\cite{exponents}, that $\mu = \nu$, it follows that the dynamical exponent z = 2.0 .

The conductivity in the insulating phase, normalized to the critical 
conductivity, is plotted in Fig. 3 on a semilogarithmic scale as a function 
of $(T^*/T)^{1/2}$, where $T^* \propto (\Delta S)^y$. For $T^*/T>10$, the 
conductivity obeys the exponentially-activated hopping form predicted by Efros and Shklovskii\cite
{efrosshklovskii} in the presence of a gap in the density of states due to 
electron-electron interactions, with a temperature-dependent prefactor given 
by the critical curve, namely:
\begin{equation}
\sigma (T) \propto T^{1/2} exp [(T^*/T)^{1/2}].
\label{equation}
\end{equation}
Deviations are evident for $T^*/T<10$.  
In this regime hopping energies are comparable or larger than the 
energy width of the Coulomb gap, and a crossover has been suggested and 
observed to Mott variable-range-
hopping 
with an exponent $1/4$ rather than $1/2$.  The inset, which shows normalized 
conductivity as a function of $(T^*/T)^{1/4}$, 
demonstrates that Mott hopping is not observed in uncompensated Si:B for any range 
of $T^*/T$.  

We now examine the behavior of the conductivity on the metallic side.  Eq. (1) 
can be rewritten as:
\begin{equation}
\sigma (t, T) = (S-S_c)^\mu f'[T/(S-S_c)^{z\nu}].
\label{eq}
\end{equation}
with a different universal function $f'$.  The ratio $\sigma (t, T)/(\Delta S/S_c)^\mu$ is 
shown as a function of $T^{1/2}/(\Delta S/S_c)^{z\nu/2}$ in Fig. 4.  The conductivity 
is everywhere consistent with the form calculated in 
the weak disorder regime (where perturbative calculations are valid)\cite{altandaron}, namely:
\begin{equation}
\sigma (T) = \sigma (0) + BT^{1/2}.
\label{eq}
\end{equation}
with $\sigma(0) \propto (S-S_c)^\mu$.

The critical conductivity exponent $\mu = 1.6$ found in our experiments 
is considerably larger than other determinations, which range between 
0.5\cite{mikko,haller} and (at most) 1.3\cite{stupp} in uncompensated 
doped semiconductors
\cite{mottsarachik}.  
We caution that a direct comparison may not be warranted for several reasons.  
Given the sensitivity of Si:B to uniaxial stress, it is possible that small 
stress variations result in an inhomogeneous stress distribution and a 
consequent averaging over a sample consisting of portions that are at different 
``distances" t from the transition  This could ``smear" the transition and 
yield a large value of $\mu$; however, if inhomogeneities were sufficiently 
serious to cause a measurable increase in $\mu$, they would probably cause 
measurable deviations from scaling as well.  It is important to note also that the 
temperature dependence of stressed and unstressed samples of comparable 
conductivities are unambiguously different\cite{bogdanovich}.  
This suggests that the question of ``universality" of the critical exponent
obtained using stress
or dopant concentration to tune the transition needs to be examined in more detail.

	It has been suggested recently\cite{Zimanyi} that disordered systems
may violate the Chayes et al.\cite{Chayes} inequality $ \nu \geq 2/d$ 
(as some of these uncompensated semiconductors appear to); self averaging 
breaks down for such sytems.  The system would then become
inhomogeneous as the critical point is approached, which
could imply that the transition is ultimately of the percolation type.
 The conductivity exponent 1.6 obtained in our 
experiments is close to that expected for classical percolation in three 
dimensions; however, these ideas need to be examined in more detail, and other 
factors must be ruled out. 

	To summarize, scaling provides an excellent description of the 
conductivity near the metal-insulator 
transition in uniaxially stressed Si:B.  Based on data at many values of 
stress and temperature, the scaling functions in the insulating and 
metallic phases yield particularly reliable 
determinations of the conductivity both within and outside the critical region.  
Although comparison 
between different systems continues to be problematical, we have shown for the 
first time that full temperature-stress scaling on both sides gives internally 
consistent results for the metal-insulator transition in a doped semiconductor.

	We are grateful to D. Simonian and S. V. Kravchenko for their
participation in some phases of these experiments.  We acknowledge 
valuable experimental contributions by A. Diop and L. Walkowicz.  
Our heartfelt 
thanks go to G. A. Thomas for his generous support and expert advice, help and 
interest throughout this project.  We thank T. A. Rosenbaum, M. 
Paalanen, E. Smith and S. Han for valuable experimental advice and the loan of equipment, and F. 
Pollak for useful suggestions 
and some samples.  M. P. S. thanks G. Kotliar and D. 
Belitz for numerous discussions.  This work was supported by the 
US Department of Energy Grant No.~DE-FG02-84ER45153.  R. N. B. was 
supported by NSF grant No. DMR-9400362.

\begin{figure}
\caption{Resistivity versus temperature on a log-log scale for different values of stress.}
\label{fig1}
\end{figure}
\begin{figure}
\caption{$\sigma /\sigma_c$ versus the scaling variable $(\Delta S/S_c)/T^y$ on a log-log 
scale, with $y=1/(\nu z)=0.31.$}
\label{fig2}
\end{figure}
\begin{figure}
\caption{For the insulating phase, $\sigma /\sigma_c$ versus $(T^*/T)^{1/2}$ on a semilogarithmic 
scale.  The inset shows 
$\sigma /\sigma_c$ versus $(T^*/T)^{1/4}$.}
\label{fig3}
\end{figure}
\begin{figure}
\caption{For the metallic phase, $\sigma /[(\Delta S)/S_c]^\mu$ versus $[T/(\Delta S
/S_c)^{z \nu}]^{1/2}$.}
\label{fig4}
\end{figure}

\begin{references}
\bibitem{dynamical} This has sometimes been referred to as dynamical scaling 
because of the 
appearance of the dynamical exponent $z$; we refer to it as 
temperature-stress scaling, reserving the term ``dynamical" 
to describe scaling of the finite-frequency conductivity.
\bibitem{analysis}The analysis presented in this paper is in terms of the 
general scaling form Eq. 1,
with exponents $x$ and $y$; only in the identification of these exponents
with the conventional exponents $\nu$ and $z$ will there be the additional
implicit assumptions concerning the scaling of the diverging time scale
with temperature.
\bibitem{parallel} In these materials, the conductivity plotted as a function 
of square root of temperature gives a family of parallel lines for different
dopant concentrations in the metallic phase, and thus clearly obey scaling
with temperature and concentration in the metal.
\bibitem{exponents}Controversy regarding this equality exists 
only when there is no symmetry breaking field (pure potential
scattering in the absence of a magnetic field), and the renormalization
group flows scale to strong coupling.
\end{references}
\end {document}
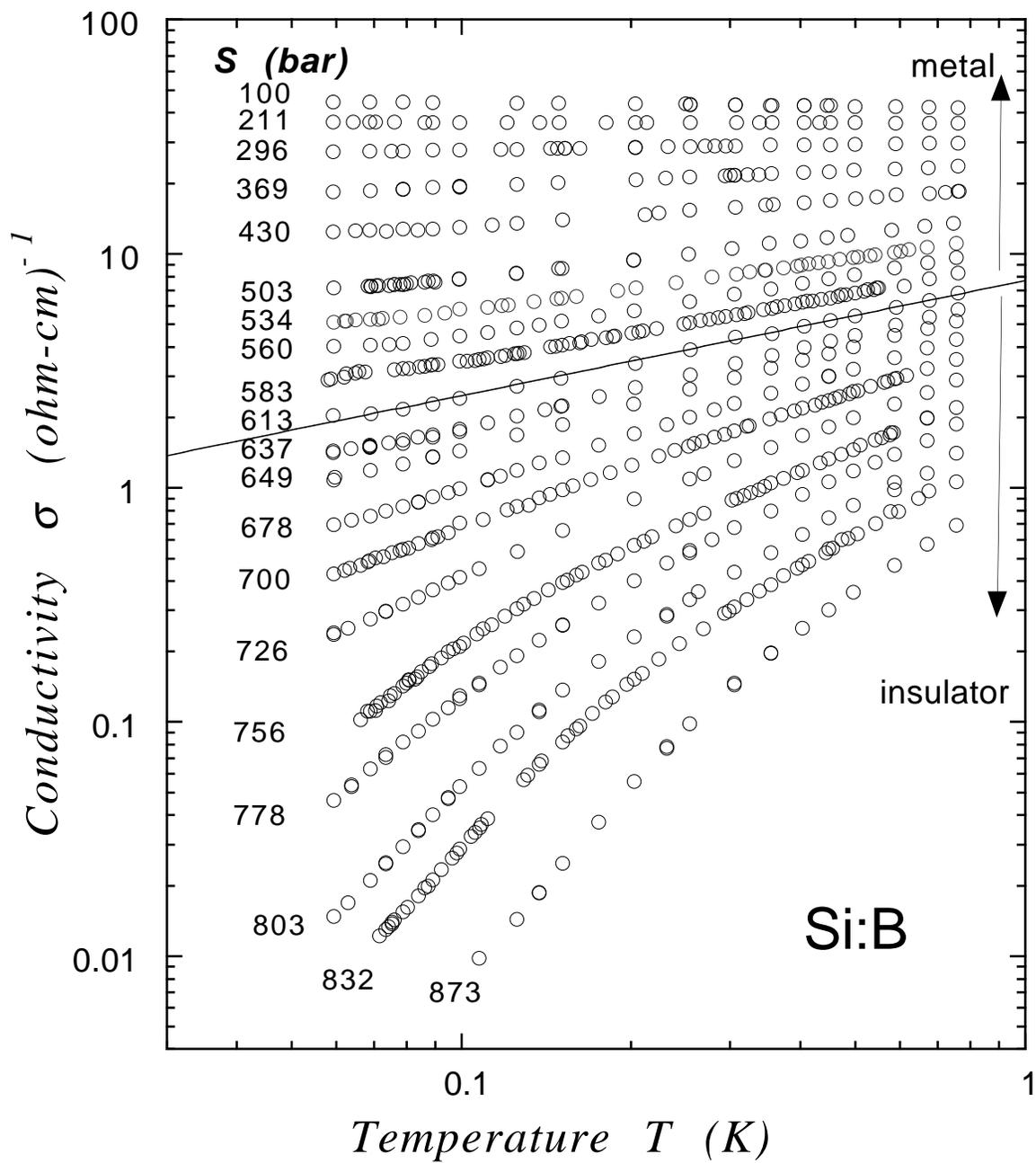
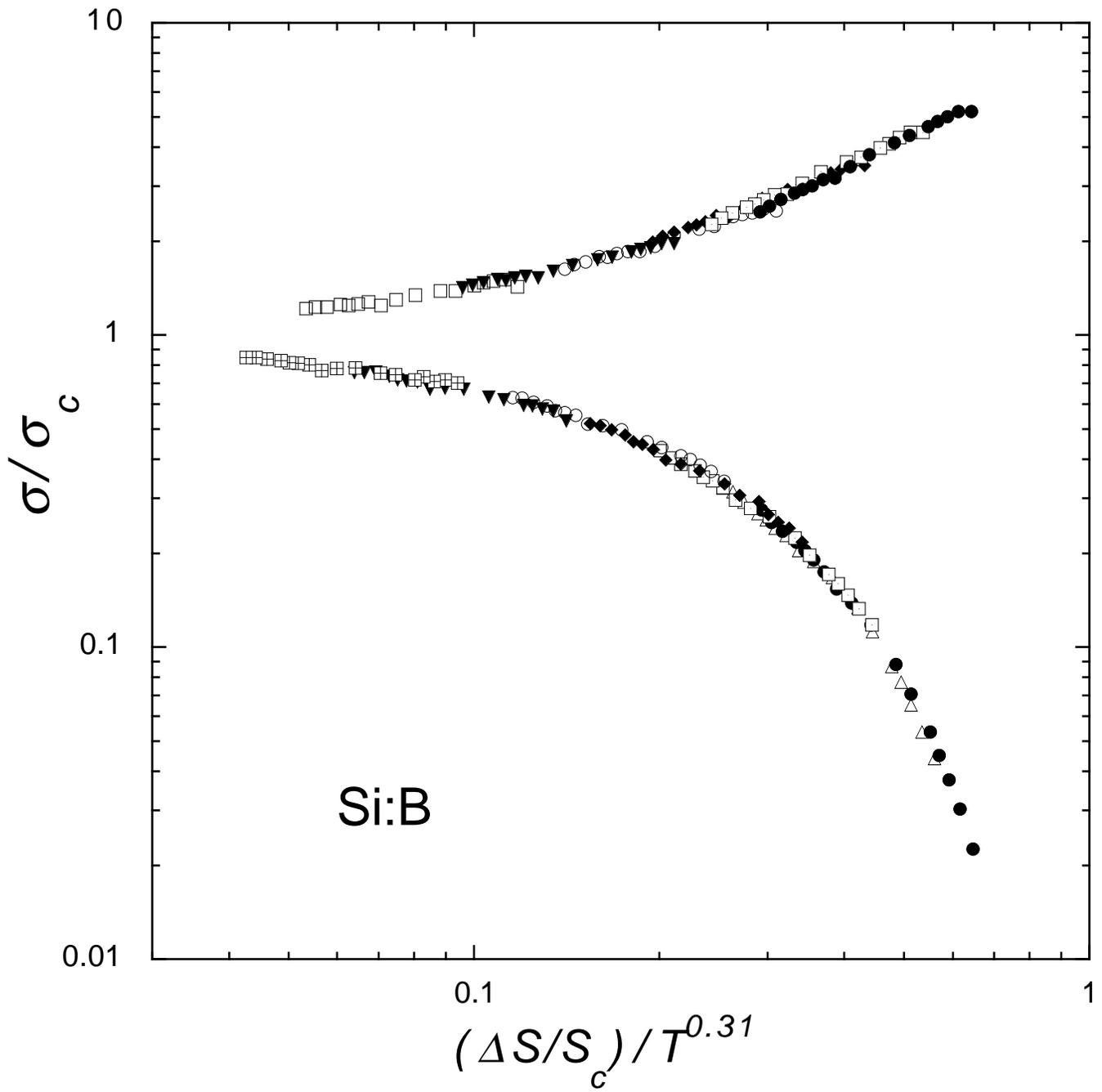
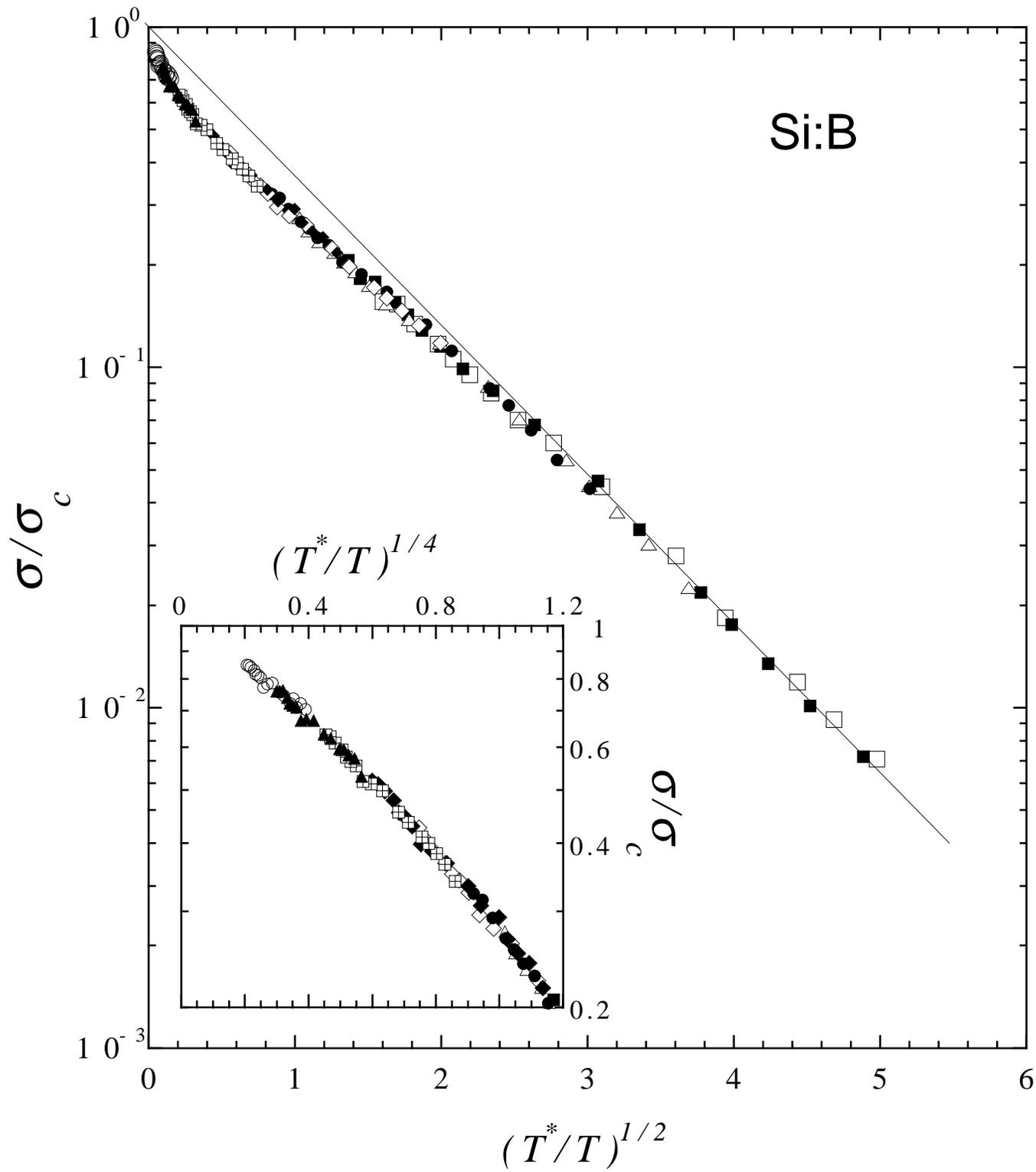
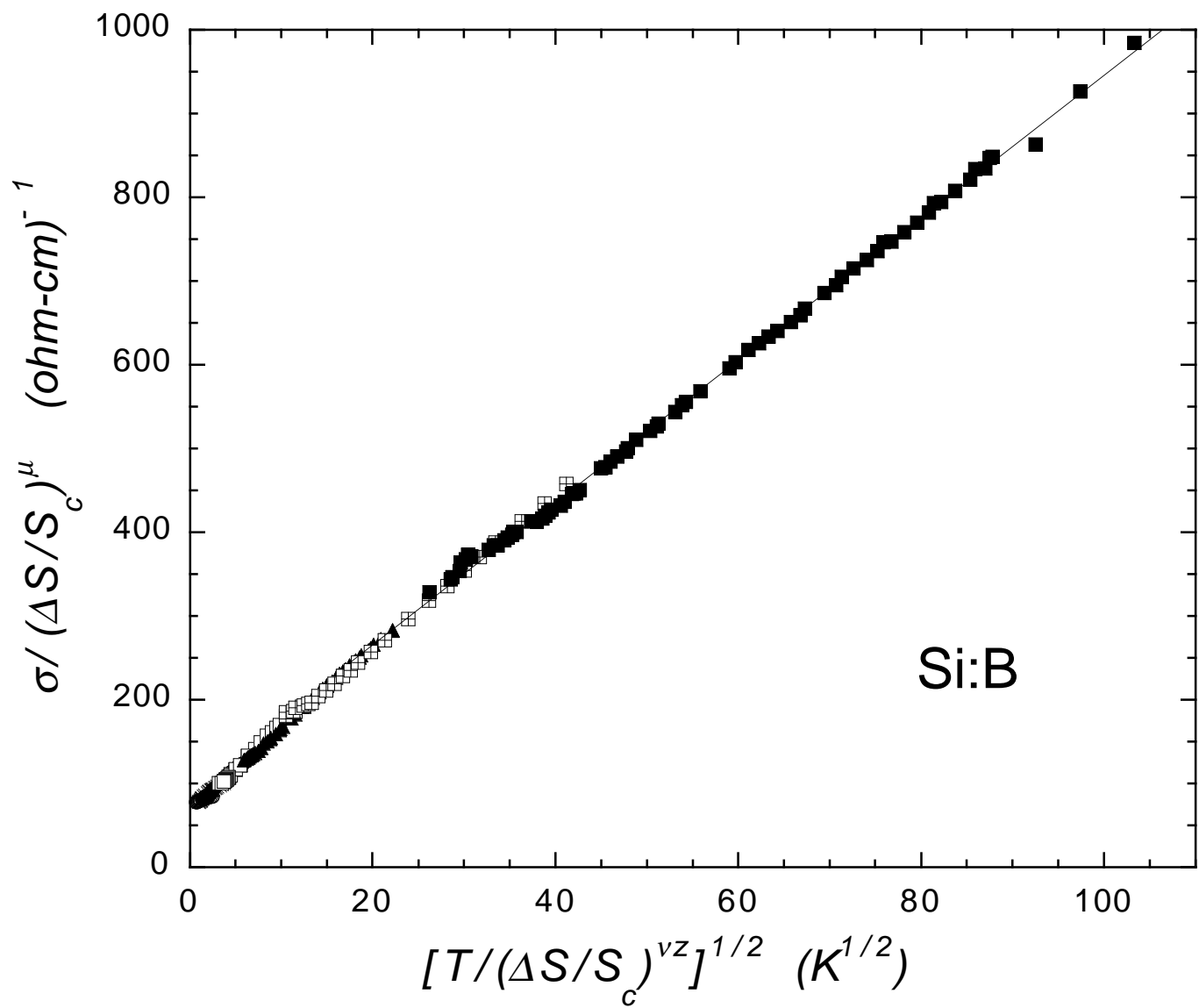
